\documentclass[showpacs,amsmath,
twocolumn,
aps,prb]{revtex4}
\usepackage{graphicx}
\usepackage{dcolumn}

\begin{document}

\title{Spin dynamics in electrochemically charged CdSe quantum dots}
\author{N. P. Stern}
\author{M. Poggio}
\author{M. H. Bartl}
\author{E. L. Hu}
\author{G. D. Stucky}
\author{D. D. Awschalom}
\affiliation{Center for Spintronics and Quantum Computation,
University of California, Santa Barbara, CA 93106}
\date{\today}

\begin{abstract}

We use time-resolved Faraday rotation to measure coherent spin
dynamics in colloidal CdSe quantum dots charged in an
electrochemical cell at room temperature.  Filling of the
1S$_{\text{e}}$ electron level is demonstrated by the bleaching of
the 1S$_{\text{e}}$-1S$_{3/2}$ absorption peak.  One of the two
Land\'{e} \textit{g}-factors observed in uncharged quantum dots
disappears upon filling of the 1S$_{\text{e}}$ electron state.  The
transverse spin coherence time, which is over 1 ns and is limited by
inhomogeneous dephasing, also appears to increase with charging
voltage.  The amplitude of the spin precession signal peaks near the
half-filling potential.   Its evolution at charging potentials
without any observable bleaching of the 1S$_{\text{e}}$-1S$_{3/2}$
transition suggests that the spin dynamics are influenced by
low-energy surface states.

\end{abstract}

\maketitle

In recent years, chemically synthesized semiconductor quantum dots
(QDs) have been used in a wide variety of applications from
QD-lasers\cite{Klimov:2000, Eisler:2002, Kazes:2002, Cha:2003} and
light emitting devices\cite{Schlamp:1997, Mattoussi:1998} to
fluorescent labels.\cite{Bruchez:1998, Chan:1998}  The size-tunable
spectrum of energy levels and the ease of colloidal nanocrystal
synthesis make these nanocrystals versatile systems for the study of
confined charges; the ability to optically excite carriers has
spurred work on coherent spin dynamics in colloidal QDs revealing
multiple Land\'{e} \textit{g}-factors and spin coherence times
reaching several nanoseconds at room temperature.\cite{Gupta:1999,
Gupta:2002} Advances in  electrochemical charging of such
nanocrystals have given researchers exquisite control over the
electronic state of the QD and have opened new avenues in the study
of state-filling effects in colloidal QDs.\cite{Wang:2001,
Guyot-Sionnest:2003, Wang:2004} In this paper we study the effects
of electrochemical charging on the spin dynamics of optically
excited carriers in an ensemble of CdSe QDs.  The filling of QD
energy levels affects properties such as the transverse spin
lifetime and the Larmor precession frequencies, revealing the
dependence of QD spin coherence on surface states and the externally
controlled electron occupation.

Stearic acid capped 6-nm diameter CdSe nanocrystal QDs are
synthesized and purified by standard colloidal chemistry
methods\cite{Qu:2001} and are dissolved in a  9:1 (by volume)
hexane:octane mixture. Following the methods of Guyot-Sionnest and
co-workers,\cite{Guyot-Sionnest:2003, Wang:2004} close-packed CdSe
nanocrystal films are formed by drop-casting the QDs onto a quartz
slide covered with a transparent indium-tin-oxide (ITO) electrode
modified with 4-aminobutyl-dimethylmethoxysilane linker molecules.
Dipping these films into a 5mM solution of $1,7$-diaminoheptane in
anhydrous methanol followed by a bake for 3 hours at 70$^\circ$C
cross-links the 'loosely' packed QDs into tightly close-packed
arrays.  The nanocrystal-coated slide is then immersed in a $0.1$ M
electrolyte solution of tetrabutylammonium tetrafluoroborate
dissolved in anhydrous N,N-dimethylformamide inside a fluorometer
cell.  A Pt counter electrode and a Ag pseudo-reference electrode
complete a three-terminal electrochemical cell with the ITO working
electrode.  The preparation of the QD film and assembly of the cell
are conducted inside of an argon atmosphere glove box. Before
removal from the glove box, the cell is sealed with epoxy and
enclosed within a vacuum-tight vessel to prevent oxidation; external
feedthroughs connect the contacts within the vessel to a
potentiostat which maintains a voltage $V_{\text{cell}}$ between the
reference and working electrodes.  Negative $V_{\text{cell}}$
injects electrons into the 1S$_{\text{e}}$ state of the QDs,
evidenced by bleaching of the 1S$_{\text{e}}$-1S$_{3/2}$ interband
absorbance peak measured using a white light source and a
spectrometer.\cite{Wang:2004}  After measurement at the desired
$V_{\text{cell}}$, a positive bias of $V_{\text{cell}} = 0.5$ V is
applied in order to discharge the quantum-confined states and any
long-lived localized trap states which may also have been filled.
Repeated charging can cause the bleaching process to slow or
eventually cease as oxygen or other impurities induce irreversible
chemical reactions in the film and electrolyte. Careful fabrication
of the cross-linked QD film and assembly of the air-tight
electrochemical cell are critical for a sample with long-lived
reproducible charging behavior.

Typical optical absorbance spectra $\alpha(E)$ are shown in Fig. 1b,
showing the 1S$_{\text{e}}$-1S$_{3/2}$ absorption peak bleaching
with negative $V_{\text{cell}}$.  The bleaching spectrum is
calculated as $\alpha - \alpha_{0} / \alpha_{0}$, where $\alpha$ is
the film's absorption at a given $V_{\text{cell}}$ and $\alpha_{0}$
is the absorption for $V_{\text{cell}} = 0$ V, and is shown in Fig.
1c for representative values of $V_{\text{cell}}$. The average
number of electrons occupying the 1S$_{\text{e}}$ state in the QD
ensemble can be controlled between 0 and 2 with the appropriate
$V_{\text{cell}}$ so that the degree of bleaching has a direct
relation to the occupation of the 1S$_{\text{e}}$
state.\cite{Wang:2004} $V_{\text{half}}$ corresponds to an average
electron occupation of 1 in the 1S$_{\text{e}}$ state of each
quantum dot and $\alpha - \alpha_{0} / \alpha_{0} = 0.5$.
$V_{\text{half}}$ varies from $-0.85$ to $-1.0$ V between QDs with
nominally similar sizes due to variations in cell assembly and film
quality. Bleaching of the 1S$_{\text{e}}$-1S$_{3/2}$ transition
occurs over a range of $\pm 0.1$ V around $V_{\text{half}}$.

\begin{figure}\includegraphics{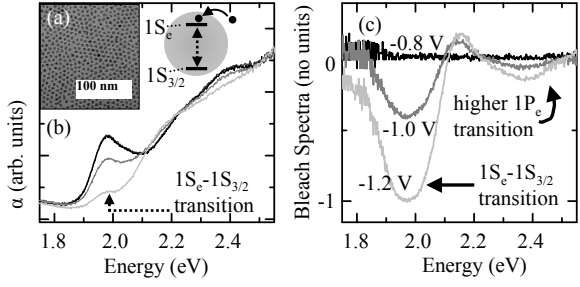}\caption{\label{fig1}
(a) TEM image of a close-packed CdSe nanocrystal film. (b)
Absorbance spectra $\alpha(E)$ at $V_{\text{cell}} = -0.8$ V(black),
$-1.0$ V (gray), and $-1.2$ V (light gray) along with a schematic of
relevant energy levels in charging experiments.  $V_{\text{cell}} =
0.0$ V is identical to $V_{\text{cell}} = -0.8$ V. (c)  Bleaching
spectra relative to $V_{\text{cell}} = 0.0$ V as described in the
text.  Bleaching of the 1P$_{\text{e}}$ electron level can be seen
at $2.35$ eV. }\end{figure}

Coherent spin dynamics are measured by time-resolved Faraday
rotation (FR) in the Voigt geometry.\cite{Crooker:1997} A
mode-locked 76-MHz Ti:Sapphire laser pumps an optical parametric
oscillator, producing 200-fs pulses tunable between energies of 1.97
and 2.21 eV which are split into a circularly polarized 1-mW pump
and linearly polarized 30-$\mu$W probe beam.  We also use a
regeneratively amplified Ti:Sapphire laser to pump two optical
parametric amplifiers (OPA) allowing independent tuning of pump and
probe energies.  The dual OPA system improves signal-to-noise, but
the qualitative features of the FR were unchanged between laser
systems.  The pump helicity is varied at 40 kHz by a photo-elastic
modulator, while the probe beam is chopped at 1 kHz for lock-in
detection.  Both beams are focused to an overlapping 100-$\mu$m
diameter spot on the QD film which is positioned between the poles
of two permanent magnets.  The pump pulse excites spin-polarized
electron-hole pairs at time $t = 0$, which we assume relax to the
lowest energy exciton state within a picosecond.\cite{Woggon:1996,
Klimov:2000, Berezovsky:2005} The pump-excited spins precess about
the applied magnetic field $B$ at the Larmor frequency $\nu_{L} = g
\mu_{B} B / h$, where $g$ is the Land\'{e} \textit{g}-factor,
$\mu_{B}$ is the Bohr magneton, and $h$ is Planck's constant.  By
the Faraday effect, the probe beam polarization axis rotates by an
angle $\theta_{F}$ proportional to the spin polarization along the
beam path; this angle is measured as a function of the time delay
$t$ between pump and probe pulses.  For consistency, all reported FR
measurements are at room temperature with $B = 0.28$ T from a single
film with $V_{\text{half}} = -0.95$ V. Variations in charging
behavior between different samples due to nanocrystal size variation
and sample quality make direct quantitative comparison between any
two samples difficult, although the qualitative results from all
samples measured agree with those reported here.

Figure 2 shows typical time-resolved FR time scans (Figs 2a-d) and
the corresponding Fourier transforms (Figs. 2e-h) for representative
charging voltages with the pump and probe energies tuned to the
1S$_{\text{e}}$-1S$_{3/2}$ exciton resonance in this sample at 2.03
eV. The data for $V_{\text{cell}} = 0.0$ V are characterized by an
oscillating component superimposed on a decaying background. The
Fourier transform for this voltage (Fig. 2e), reveals two distinct
frequencies ($\nu_1$, $\nu_2$) contributing to the oscillation,
consistent with previous FR studies of CdSe QDs.\cite{Gupta:2002,
Berezovsky:2005} On charging, the FR amplitude increases while the
relative weight of the higher frequency component ($\nu_{2}$)
decreases until disappearing completely around $V_{\text{half}}$.
(Figs. 2f-h) The amplitude of the lower frequency component
($\nu_{1}$) decreases beyond $V_{\text{half}}$. Monitoring the
absorbance of the probe beam, $\alpha_{\text{probe}}$, ensures that
the QDs within  the laser focus spot are charging with the
application of $V_{\text{cell}}$ as the absorbance peak bleaches
(Fig. 3a).

\begin{figure}\includegraphics{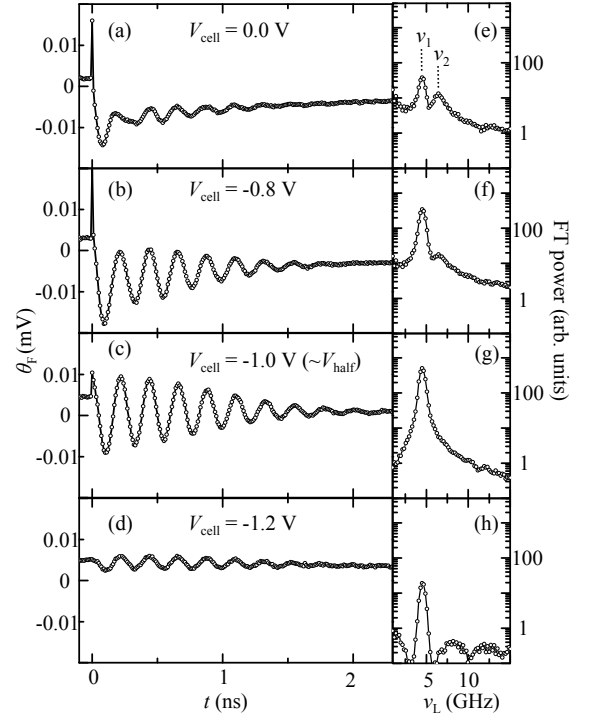}\caption{\label{fig2} (a) - (d) FR time-domain scans for representative voltages. Voltages not plotted represent a smooth evolution between the displayed results.  The fast Fourier transforms in (e) - (h), which have been smoothed for clarity, show two frequencies in uncharged nanocrystals evolving to a single precession frequency in the charged ensemble. }\end{figure}

In order to more carefully examine these parameters, the FR
time-domain data are fit to two decaying cosinusoidal components,
given by Eq. (\ref{eq1}):
\begin{eqnarray}
\label{eq1} \theta_F = \theta_0 + \theta_1
\exp(-t/T^{*}_{2,1})\cos(2\pi \nu_1 t + \phi_1)+ \nonumber\\
\theta_2\exp(-t/T^{*}_{2,2})\cos(2\pi \nu_2 t + \phi_2) +
\theta_{exp}\exp(-t/\Delta_t)
\end{eqnarray}
where $\theta_{0}$ is an offset, $\theta_{1} (\theta_{2})$ the
amplitude of the first (second) frequency component, $T^{*}_{2,1}
(T^{*}_{2,2})$ the transverse spin coherence time, $\nu_1$ $(\nu_2)$
the Larmor frequency, and $\phi_1 (\phi_2)$ the phase,
$\theta_{exp}$ and $t$ the amplitude and decay time respectively of
the non-oscillating exponential background.  At voltages
$V_{\text{cell}} < -0.9$ V, the second frequency component ($\nu_2$)
is not present and only one of the cosinusoidal terms is kept in the
fit function.

The non-oscillating component with a decay time of $t = 0.9 \pm 0.1$
ns disappears when the film is charged beyond $V_{\text{half}}$.
(Fig. 3b) The origin of this component is unclear, as earlier
studies of confined nanostructures have attributed similar
non-oscillating signals as to the decay of hole spins pinned along
the growth axis,\cite{Crooker:1997} the leakage of longitudinal spin
relaxation signal,\cite{Gupta:2002} surface carrier
trapping,\cite{Gupta:1999} or the decay of non-oscillating exciton
populations.\cite{Tartakovskii:2004}  The quenching of
$\theta_{\text{exp}}$ with charging is consistent with the latter
two explanations.

\begin{figure}\includegraphics{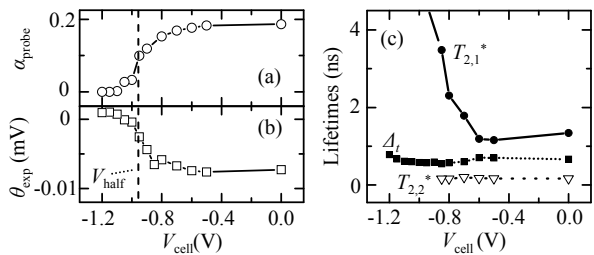}\caption{\label{fig3}
(a) Absorbance of the probe beam, measuring the local bleaching at
the laser focus spot.  (b) Amplitude of the non-oscillating
component $\theta_{\text{exp}}$, which vanishes on charging. (c) Fit
results for the spin coherence times $T^{*}_{2,1}$ (circles),
$T^{*}_{2,2}$ (triangles), and the Gaussian dephasing time
$\Delta_t$ (squares).  For $T^{*}_{2,1} \gg \Delta_t$, the fitting
algorithm cannot quantitatively distinguish between a purely
Gaussian lifetime and the exponential lifetime $T^{*}_{2,1}$,
suggesting that the spin relaxation is completely dominated by
inhomogeneous dephasing for $|V_{\text{cell} }|> 1.0$ V.
}\end{figure}

The lower frequency FR amplitude $\nu_1$ decays with a Gaussian
envelope, indicating that the relaxation of coherent spins in the QD
ensemble is dominated by inhomogeneous dephasing.  The effect of the
inhomogeneous QD size can be modeled as a Gaussian distribution of
\textit{g}-factors with width $\Delta_g$. Averaging the FR
contributions over the ensemble introduces an additional Gaussian
envelope to the oscillatory terms of Eq. 1 with a width $\Delta_t =
\hbar / \mu_B B \Delta_g$.\cite{Gupta:2002} Fits to the FR yield
both the Gaussian decay width and the transverse spin lifetimes
(Fig. 3c).  Since $\Delta_t$ is about $0.7$ ns for all
$V_{\text{cell}}$, the dephasing is due to structural properties
such as orientation and size variation that are not affected by
changes in electronic occupation.  Assuming all inhomogeneities are
size related, this dephasing rate corresponds to a \textit{g}-factor
width $\Delta_g =0.06$ and a size distribution of about 15\%. The
size distribution of the synthesized QDs is estimated to be closer
to 5\%, suggesting contributions to the inhomogeneous broadening by
shape anisotropy and random nanocrystal orientation which are not
accounted for in this simple size-averaging procedure. The spin
coherence time of the $\nu_1$ component, $T^{*}_{2,1}$, increases
dramatically with charging, though quantitative fit results for
coherence times above $2$ ns are not reliable because the vanishing
of the FR signal due to dephasing.   This result is consistent with
the large increase in spin lifetimes reported in \textit{n}-doped
self-assembled InAs-GaAs QDs compared to undoped
QDs.\cite{Cortez:2002}  Further experiments in systems with less
inhomogeneity or on single quantum dots will help avoid the
limitations of inhomogeneous broadening and isolate the effects of
charging upon the spin lifetime.

The amplitudes of the two FR components behave quite differently
from each other (Fig. 4a). The lower frequency amplitude
$\theta_{1}$ increases significantly, reaching a maximum near
$V_{\text{half}}$ before decreasing toward zero in doubly-charged
QDs.  $\theta_{2}$ vanishes sharply at $V_{\text{half}}$ due to the
presence of charges in the 1S$_{\text{e}}$ state.

Comparison between Figs. 3a and 4a show that $\nu_{1}$ increases at
$V_{\text{cell}}$ well below the 1S$_{\text{e}}$ charging near
$V_{\text{half}}$.  The amplitude increase between 0 V and
$V_{\text{half}}$ may be due to the filling of low-energy surface
states on the nanocrystals which have been shown to provide fast
non-radiative pathways for carrier relaxation. Surface defect states
in uncharged QDs, in addition to influencing energy
levels\cite{Franschetti:2000} and photoluminescence
dynamics\cite{Wang:2004}, account for the relaxation of up to half
of optically injected carriers within a few ps (Fig.
4c).\cite{Klimov:1999, Klimov:2000}  For $V_{\text{cell}}$ between 0
V and $V_{\text{half}}$, these empty surface states in the QD band
gap can be filled by electrochemically injected charges, blocking
the relaxation pathway and increasing the number of QDs in the
ensemble with long-lived coherent electron populations (Fig. 4d).
This interpretation is supported by the disappearance in charged QDs
of the spikes in the FR near $t=0$ which may denote decay processes
on the order of a few ps (Fig. 2a-d).  The FR amplitude reduction
for charging beyond $V_{\text{half}}$ is due to Pauli blocking of
absorptive 1S$_{\text{e}}$-1S$_{3/2}$ transitions, as evidenced by
the commensurate bleaching (Fig. 1c) and reduction in probe beam
absorption (Fig. 3a) in this regime.

\begin{figure}\includegraphics{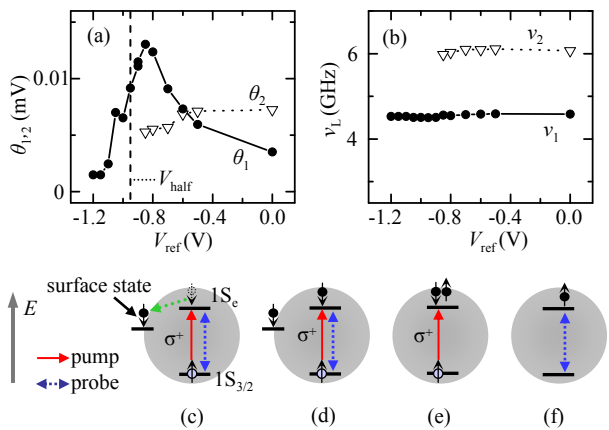}\caption{\label{fig4} (Color)
(a) FR amplitudes $\theta_1$ and $\theta_2$ as functions of
$V_{\text{cell}}$. (b) Larmor precession frequencies $\nu_1$ and
$\nu_2$ corresponding to g-factors of $g_1 = 1.15$ and $g_2 = 1.63$.
(c)-(f) Schematic diagrams of charged QD spin dynamics under
$\sigma^{+}$ pump excitation. For $\sigma^{-}$ pump pulses, all spin
directions should be reversed. (c) In a fraction of CdSe QDs,
optically injected electrons relax non-radiatively (green arrow)
into surface defect states within a few ps, leaving no
1S$_{\text{e}}$ electron to probe. (d) For a $V_{\text{cell}}$ that
fills the surface states, optically injected electrons remain in the
1S$_{\text{e}}$ level and FR can be measured in all QDs. (e) A
circularly polarized pump pulse can only create an exciton in a
charged QD with one spin orientation of the electrically injected
charge. The singlet in the 1S$_{\text{e}}$ state prevents probe
pulses from interacting with this QD. (f) Probe pulses can measure a
singly-charged QD which has not absorbed a pump pulse; there is no
exciton precession, however. }\end{figure}

The two constant precession frequencies (Fig. 4b) correspond to
Land\'{e} \textit{g}-factors of $g_{1} = 1.15$ and $g_{2} = 1.63$,
in agreement with what would be expected for a 6-nm diameter
QD.\cite{Gupta:1999} The origin of the two \textit{g}-factors in
CdSe QDs has been attributed to multiple sources in the literature.
\cite{Schrier:2003} Theoretical calculations from effective mass
models suggest that the lower frequency $g_{1}$ is electron spin
precession.\cite{Rodina:2003,Gupta:2002}  Ref.
\onlinecite{Gupta:2002} follows this assignment and subsequently
attributes $g_2$ to exciton precession in a subset of quasispherical
QDs where the spin-splitting due to shape and wurtzite crystal
anisotropies cancel.  Another interpretation of the time-resolved FR
data in Ref. \onlinecite{Gupta:2002} proposed by Chen and Whaley
\cite{Chen:2004} attributes the two observed \textit{g}-factors to
the two anisotropic components of the \textit{g}-tensor $g_{\perp}$
and $g_{z}$ predicted by tight-binding calculations.   As noted in
Ref. \onlinecite{Chen:2004}, however, this calculation does not
account for the random orientation of the nanocrystal axes relative
to the spin injection axis.  Previous experimental work shows that
anisotropic \textit{g}-factors should lead to a single averaged
frequency in the FR data.\cite{Salis:2001} Further, the markedly
different qualitative behavior of the two precession frequencies
with charging in the current experiment supports the interpretation
by Ref. \onlinecite{Gupta:2002} of distinct states giving rise to
the components.  These details favor the conclusion that $g_{2}$ is
an excitonic \textit{g}-factor, but neither the original FR
measurements nor the charging experiments give any details as to
which fine-structure exciton state may be involved in the
precession.

The quenching of the $g_{2}$ component can be explained within this
exciton precession model by noting that in an idealized QD,
selection rules dictate that circularly polarized pump pulses can
only interact with a subset of singly-charged QDs in the ensemble.
When the single 1S$_{\text{e}}$ electron is spin up (down), only
$\sigma^{+}( \sigma^{-})$ polarized pump pulses can be
absorbed.\cite{Tartakovskii:2004} Since the 1S$_{\text{e}}$
electrons of the resulting charged exciton must be in a singlet
state, the delayed probe pulse cannot interact with the pumped QDs
(Fig. 4e). Thus, the FR signal from a charged QD ensemble is likely
due to QDs which had no initial charge and were optically excited,
or singly-charged QDs which have not absorbed the pump pulse because
the 1S$_{\text{e}}$ charge has the opposite spin for the pump
helicity.  For a $V_{\text{cell}}$ where QDs are at least
singly-charged, the probe pulse will not measure any QDs with
optically generated excitons, but rather only those with a single
electrically injected 1S$_{\text{e}}$ electron (Fig. 4f). Hence, for
$V_{\text{cell}}$ corresponding to a charged ensemble, only electron
spin precession is observed whereas exciton precession is quenched.

In summary, we have studied the effects of electrochemical charging
on the spin dynamics of optically injected carriers in colloidal
CdSe quantum dots.  The charging of the nanocrystals suppresses the
higher of two spin precession frequencies, which is most likely of
excitonic nature.   Further, the amplitude of the spin precession
increases up to near the half-filling potential and decreases in
completely charged dots, likely due to surface defects and Pauli
blocking.  Inhomogeneous dephasing limits interpretation of spin
coherence times, though there is evidence for an increase in
transverse electron spin lifetime.

We thank J. Berezovsky for illuminating discussions.  This work was
supported by DARPA and the NSF.  N. P. S. acknowledges the Fannie
and John Hertz Foundation.

\end{document}